# Assessments of the energy, mass and size of the Chicxulub Impactor

Hector Javier Durand-Manterola and Guadalupe Cordero-Tercero

Departamento de Ciencias Espaciales, Instituto de Geofísica, Universidad Nacional Autonoma de México.

hdurand_manterola@yahoo.com
guadalupecordero@hotmail.com

Abstract

In 1980, Alvarez and colleagues proposed that, in the transition from the Cretaceous to Paleogene, a large impactor collided with Earth being the cause of the mass extinction occurred at the limit K / Pg. In 1980 there was no known impact structure, which could be responsible for this extinction.

It was not until 1991 that an international group of researchers proposed that a circular structure between 180 and 200 km, buried under Tertiary deposits in the Yucatan Peninsula in Mexico, was the crater formed by the impact proposed by the group of Alvarez (Hildebrand et al., 1991). It is very probable that an impact of this magnitude have had large effects on the surface and in the environment. To study these effects, it is necessary to estimate the characteristics that the impactor had. The literature often mentions the nature of the impactor, and has been proposed both an asteroid and a comet, and even a comet shower that produced periodic extinctions.

However, the physical parameters of the impactor are not limited, so the aim of this study is to estimate the most relevant features of this one such as the size, mass and kinetic energy. We found that the kinetic energy of the impactor is in the range from $1.3 \times 10^{24}$ J to $5.8 \times 10^{25}$ J. The mass is in the range of $1.0 \times 10^{15}$ kg to $4.6 \times 10^{17}$ kg. Finally, the diameter of the object is in the range of 10.6 km to 80.9 km. Based on the mass of the impactor and iridium abundance in different types of meteorites, we calculate the concentration of iridium, which should be observed in the K/Pg layer. When compared with the measurements, we concluded that the best estimation is that the impactor was a comet.

1 Introduction

When an asteroid or comet impact with the surface of a planetary body instantly releases all its kinetic energy, evaporating, pulverizing and melting rock, generating a shock wave that compresses the target and eventually excavates a cavity in the ground: the impact crater.

The impact cratering is a geological process that occurs in all bodies with solid surfaces in the solar system. The morphology of these craters changes according to the size of the excavated cavity and also depends on the characteristics of the target (gravitational acceleration, type of crustal rocks, abundance of volatiles, etc..).

In 1980, Alvarez and colleagues proposed that, in the transition from the Cretaceous to Paleogene, a large impactor collided with Earth being the cause of the mass extinction occurred at the limit K / Pg. In 1980 there was no known impact structure, which could be responsible for this extinction.

It was not until 1991 that an international group of researchers proposed that a circular structure between 180 and 200 km, buried under Tertiary deposits in the Yucatan Peninsula in Mexico, was the crater formed by the impact proposed by the group of Alvarez (Hildebrand et al., 1991). It is very probable that an impact of this magnitude have had large effects on the surface and in the environment. To study these effects, it is necessary to estimate the characteristics that the impactor had. Some authors have proposed an asteroid (Alvarez, 1983) and others a comet, and even a comet shower that produced periodic extinctions (Raup and Sepkoski, 1984; Hut et al., 1987). In this study we obtained an estimation of the most important characteristics of the impactor, which are the size, mass and kinetic energy. We found that the kinetic energy of the impactor is in the range from $1.3 \times 10^{24}$ J to $5.8 \times 10^{25}$ J. The mass is in the range of $1.0 \times 10^{15}$ kg to $4.6 \times 10^{17}$ kg. Finally, the diameter of the object is in the range of 10.6 km to 80.9 km. Based on the mass of the impactor and iridium abundance in different types of meteorites, we calculate the concentration of iridium, which should be observed in the K/Pg layer. When compared with the measurements, we concluded that the best estimation is that the impactor was a comet.

2 Impactor Energy

From the crater diameter and several models that have been published in the literature, one can give an estimate of the kinetic energy that the object had at the time of the collision. To do this we used four models, one developed by us and three mentioned in the literature (Dence et al., 1977, de Pater and Lissauer, 2001, McKinnon et al., 2003, Faure and Mensing, 2007; Melosh, 2011).

2.1 First Model

In 1977, Dence and his collaborators, using the diameters of the craters produced by atomic explosions and the energy of these, they obtained the following empirical equation that relates the diameter of the crater with the energy released:

$$D = 1.96 \times 10^{-5} E^{0.294} \quad (2.1.1)$$

Where D is in km and E in Joules. This model is the simplest of the four that we used in this work (Dence et al., 1977, Faure and Mensing 2007, p 134).

From this equation it is easy to solve the energy E as a function of crater diameter.

$$E = \left[\frac{D}{1.96 \times 10^{-5}}\right]^{3.4} \quad (2.1.2)$$

2.2 Second Model

Much later, in 2001, de Pater and Lissauer mention another model, based on laboratory experiments. This model is more complicated than the previous one, also relates crater diameter D with the kinetic energy of the impactor E, but requires other parameters including the radius of the impactor (de Pater and Lissauer, 2001, pp. 165):

$$D = 1.8\, \rho_i^{0.11} \rho_p^{-\frac{1}{3}} g_p^{-0.22} (2r)^{0.13} E^{0.22} (\text{sen}\,(\theta))^{\frac{1}{3}}$$
$$(2.2.1)$$

Where $\rho_i$ is the density of the impactor; $\rho_p$ is the density of the target; $g_p$ is the planet's surface gravity, r is the radius of the impactor and $\theta$ the angle of the trajectory of the impactor relative to the horizontal at the point of impact.

Substituting in (2.2.1) the value of r given by equation (3.4) and solving for E we have

$$E = \left[\frac{D}{1.8\, \rho_i^{0.11} \rho_p^{-\frac{1}{3}} g_p^{-0.22} \left(2\left[\frac{6}{4\pi\rho V^2}\right]^{1/3}\right)^{0.13} (\text{sen}(\theta))^{\frac{1}{3}}}\right]^{3.798} \quad (2.2.2)$$

where V is the velocity of the impactor.

2.3 Third model
The third model is a recent one (Melosh, 2011)

$$D_{tr} = 1.161 \left(\frac{\rho_i}{\rho_p}\right)^{1/3} (2r)^{0.78} V^{0.44} g^{-0.22} \text{sen}^{1/3}\theta \quad (2.3.1)$$

where r and V are the radius and the entry speed of the projectile, respectively, g is the surface gravity of the target, θ is the angle of entry. $\rho_i$ and $\rho_p$ are respectively the density of the projectile and the target.

Substituting the radius of the projectile given by (3.4) in (2.3.1) and solving the kinetic energy, we have:

$$E_c = \left[\frac{D_{tr}}{1.645}\left(\frac{\rho_p}{\rho_i}\right)^{\frac{1}{3}} \rho_i^{0.26} V^{0.08} g^{0.22} sen^{-\frac{1}{3}}\theta\right]^{\frac{1}{0.26}} \quad (2.3.2)$$

This model uses an empirical scaling law relating energy impactor with transient crater diameter, but not with the final diameter.
En 2003 McKinnon et al. In 2003 McKinnon examined three empirical scaling laws that relate the final diameter crater, D, with transient crater diameter, $D_{tr}$.

$$D = D_c^{-0.18 \pm 0.05} D_{tr}^{1.18 \pm 0.06} \quad (2.3.3)$$

$$D = 1.17\, D_c^{-0.13} D_{tr}^{1.13} \quad (2.3.4)$$

$$D = 1.02\, D_c^{-0.086} D_{tr}^{1.086} \quad (2.3.5)$$

where $D_c$ is the diameter of transition from simple to complex craters, which in the case of the Earth is between 3 and 5 km (Melosh, 1989). In this study we will take it as 4 km.
From these three relations we can solve $D_{tr}$, and we have:

$$D_{tr} = [DD_c^{0.18 \mp 0.05}]^{0.85 \mp 0.04} \quad (2.3.6)$$

$$D_{tr} = [0.855 DD_c^{0.13}]^{0.89} \quad (2.3.7)$$

$$D_{tr} = [0.98 DD_c^{0.086}]^{0.921} \quad (2.3.8)$$

We can substitute these relations into (2.3.2) and thus obtain the relation between energy and final diameter.

2.4 Fourth Model
We developed this model that relates the transient crater diameter with the energy of the impactor.

Assuming all the mass, m, excavated during transient crater formation is launched into space with escape velocity, Ve, we can estimate an upper limit of energy impactor:

$$E = \frac{1}{2}mV_e^2 \quad (2.4.1)$$

On the other hand, considering the transient crater like a half-sphere then the extracted mass to form it, is:

$$m = \frac{2}{3}\rho\pi R^3 \quad (2.4.2)$$

where ρ is the density of target rocks and R is the radius of the transient crater.
Substituting (2.4.2) into (2.4.1), we have

$$E = \frac{1}{3}\rho\pi R^3 V_e^2 \quad (2.4.3)$$

As R = $D_{tr}$/2 then

$$E = \frac{1}{24}\rho\pi D_{tr}^3 V_e^2 \quad (2.4.4)$$

This would be the energy, as function of diameter, for the extreme case the entire mass of the crater was ejected to escape velocity. This would be a maximum limit of energy for a given diameter.

An estimate of the minimum energy required to form a crater of a given diameter, is obtained if the entire mass of the crater was deposited outside this. That is, considering the energy required to move a slab of material with a width dh, from a depth h to the surface, and then integrating from the surface to the maximum depth, R. In this case the energy would be:

$$E = \int_0^R \rho\pi g h r^2 \, dh \quad (2.4.5)$$

where $r^2$ is

$$r^2 = R^2 - h^2 \quad (2.4.6)$$

Therefore substituting (2.4.6) into (2.4.5) we have

$$E = \int_0^R \rho\pi g h (R^2 - h^2) dh \quad (2.4.7)$$

And integrating

$$E = \frac{\rho\pi g R^4}{2} - \frac{1}{4}\rho\pi g R^4 = \frac{\rho\pi g R^4}{4} \quad (2.4.8)$$

As R = $D_{tr}$/2, then

$$E = \frac{\rho\pi g}{64} D_{tr}^4 \quad (2.4.9)$$

As (2.4.4) and (2.4.9) are extreme values, both unrealistic, then the real energy will be within this range. In this paper we considered the average energy as the true value.

$$E = \frac{\pi\rho}{16} D_{tr}^3 \left[\frac{V_e^2}{3} + \frac{gD_{tr}}{8}\right] \quad (2.4.10)$$

To relate the energy to the final diameter of the crater, we use equations (2.3.6) to (2.3.8).

3 Impactor mass and radius
From kinetic energy equation we can solve for the mass. As we know the energy, then we have the mass as a function of the impactor velocity.

$$m = \frac{2E}{V^2} \quad (3.1)$$

Assuming that the impactor was spherical with radius r, then we have that mass is

$$m = \frac{4}{3}\pi r^3 \rho_i \quad (3.2)$$

where $\rho_i$ is the average density of the impactor.
Substituting (3.1) in (3.2) we have

$$\frac{2E}{V^2} = \frac{4}{3}\pi r^3 \rho_i \quad (3.3)$$

Solving for r and multiplying by 2, we have the diameter of the impactor

$$d = 2\left[\frac{6E}{4\pi\rho_i V^2}\right]^{1/3} \quad (3.4)$$

Table 1
Energy, Mass and Diameter of the Chicxulub impactor according to the four models mentioned in the text.

| Model | Type of object | Energy (J) x $10^{24}$ | | Mass (kg) | | Diameter (km) | |
|---|---|---|---|---|---|---|---|
| | | Min | Max | Min | Max | Min | Max |
| 1 | Stony | 0.5 | 0.7 | $5.7 \times 10^{14}$ | $8.5 \times 10^{15}$ | 6.8 | 16.8 |
| | Metallic | | | | | 5.1 | 12.7 |
| | Comet | | | $1.8 \times 10^{14}$ | $5.3 \times 10^{15}$ | 5.9 | 18.3 |
| 2 | Stony | 3.0 | 6.6 | $5.3 \times 10^{15}$ | $5.6 \times 10^{16}$ | 14.4 | 31.6 |
| | Metallic | 2.4 | 5.3 | $4.3 \times 10^{15}$ | $4.5 \times 10^{16}$ | 10.1 | 22.1 |
| | Comet | 3.9 | 9.6 | $2.4 \times 10^{15}$ | $4.5 \times 10^{16}$ | 14.1 | 37.4 |
| 3 | Stony | 1.3 | 6.4 | $2.2 \times 10^{15}$ | $5.6 \times 10^{16}$ | 10.8 | 31.6 |
| | Metallic | 1.0 | 5.0 | $1.7 \times 10^{15}$ | $4.4 \times 10^{16}$ | 7.5 | 22.0 |
| | Comet | 1.7 | 9.4 | $1.0 \times 10^{15}$ | $4.6 \times 10^{16}$ | 10.6 | 37.6 |
| 4 | Stony | 4.4 | 58 | $5.3 \times 10^{15}$ | $7.4 \times 10^{17}$ | 14.4 | 74.5 |
| | Metallic | | | | | 10.8 | 56.0 |
| | Comet | | | $1.6 \times 10^{15}$ | $4.6 \times 10^{17}$ | 12.4 | 80.9 |

4 Results

Because the interval of time that separates us from the formation of Chicxulub is so large, the evidence that could help to reconstruct impactor features are few. The most obvious is the diameter of the crater, which is between 180 and 200 km (Schulte et al., 2010). With these extreme values and equations (2.1.2), (2.2.2), (2.3.2), (2.4.10) we can estimate the energy of the impactor.

4.1 Impactor kinetic energy

Figure 1 shows the energy intervals estimated from the various models. In model 1 (equation 2.1.2) the energy is a monotonically increasing function of the diameter. For diameters between 180 km and 200 km, the energy takes values between $4.7 \times 10^{23}$ J y $6.7 \times 10^{23}$ J (see Table 1 and Figure 1).

In the second model, to calculate the kinetic energy of the impactor, we needed crater diameter, density of the projectile, density of the target, earth's gravity and impactor velocity. We considered the density of the projectile as 1650 kg m$^{-3}$ for comets (Greenberg, 1998), 3400 kg m$^{-3}$ for stony asteroids (Wilkison and Robinson, 2000), and 8000 kg m$^{-3}$ for iron asteroids (Hills and Goda, 1993). We took the target density as 2460 kg m$^{-3}$, which is the modal density of the limestone of Yucatan (Alonzo et al., 2003), and Earth's gravity as 9.80 m s$^{-2}$ (Tholen et al., 2000). Steel (1998) obtained an estimation of the range of velocities for bodies that cross Earth's orbit. For asteroids the interval is between 12.6 km s$^{-1}$ and 40.7 km s$^{-1}$. This result is based on measurements of the velocities of the asteroids that cross Earth's orbit.

The range for comets is between 16 km s$^{-1}$ and 73 km s$^{-1}$. This result is obtained from a theoretical calculation of the expected velocity distribution of bodies that come from the Öpik-Oort cloud.

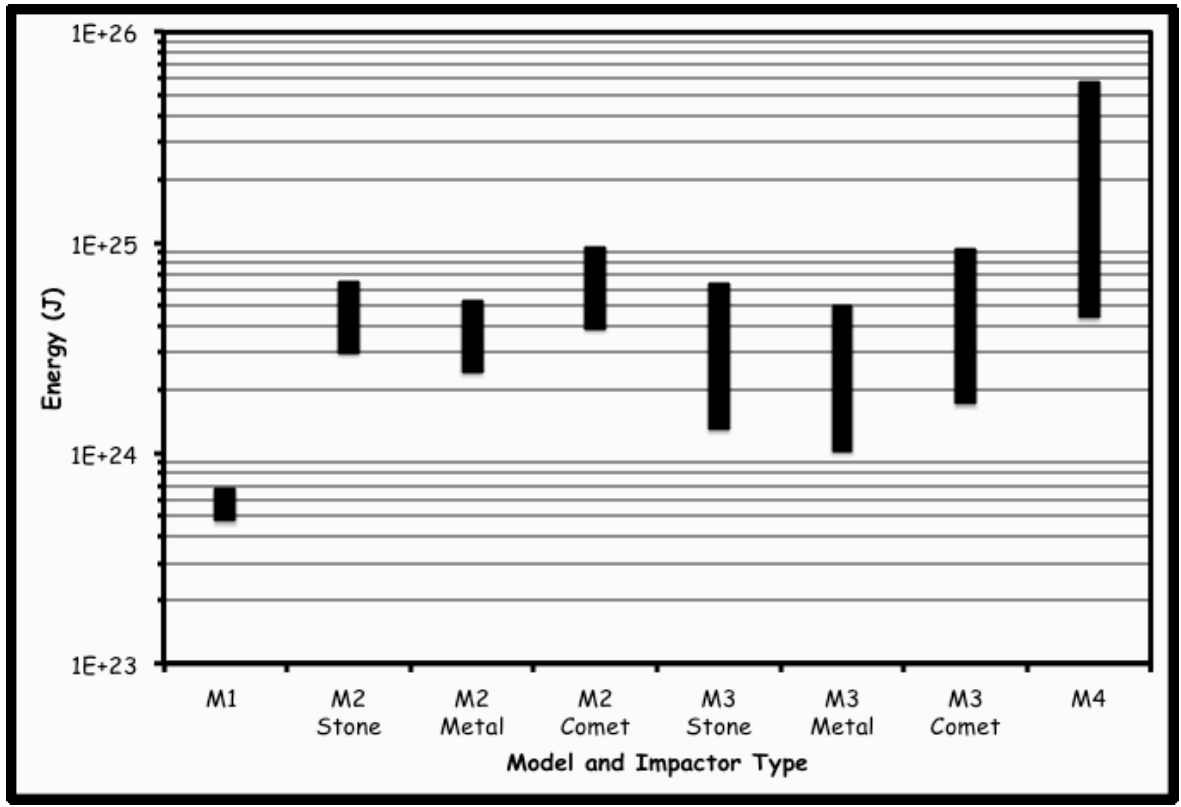

Figure 1. Energy of the impactor, according to the four models mentioned in text. Note that the values of model 1 are out range of the theoretical model 4.

Taking into account all these values and applying them to equation (2.2.2) we have that the kinetic energy, for all the objects, is between 2.4x10$^{24}$ J and 9.6x10$^{24}$ J. To see the energy for each type of impactor see Table 1 and Figure 1. To calculate the kinetic energy of the impactor, with the third model, we require the same parameters for the second model. Taking into account all these parameters and applying them to equation (2.3.2) we got that the kinetic energy is between 1.0x10$^{24}$ J and 9.4x10$^{24}$ J. To see the energy for each type of impactor see Table 1 and Figure 1.

To calculate the kinetic energy of the impactor with the fourth model, we required the crater diameter, the Earth's gravity and the escape velocity of the planet, which is 11.18 km s$^{-1}$ (Tholen et al., 2000). Applying these values to equation (2.4.10) we have that the kinetic energy is between $4.4 \times 10^{24}$ J and $5.8 \times 10^{25}$ J (Table 1 and Figure 1). This model does not distinguish between asteroids and comets.

Taking all models together, the range of values of the energy of the impactor is between $4.7 \times 10^{23}$ J and $5.8 \times 10^{25}$ J.

### 4.2 Impactor Mass

Using equation (3.1), the estimated energy for each of the models, and the velocity range estimated by Steel (1998), we calculate the impactor mass (Figure 2). Taking all models, the mass range of the impactor is between $5.7 \times 10^{14}$ kg and $4.6 \times 10^{17}$ kg (Figura 2 and Table 1).

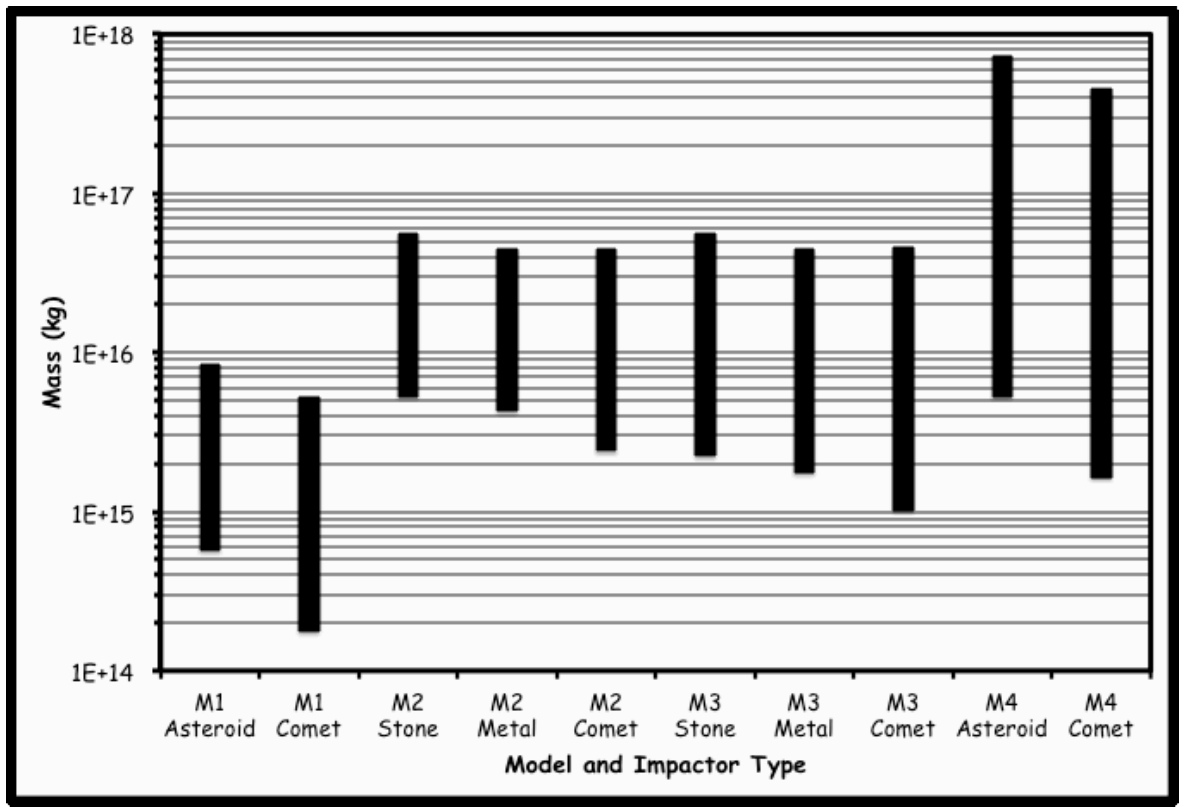

Figure 2. Mass of the impactor according to the four models mentioned in text.

### 4.3 Impactor diameter

Using equation (3.4), energy, speed and density of the impactor, we calculated the diameter, assuming that is spherical. Since in equation (3.4) the speed and density of the impactor appear, then in all models we can distinguish among comets, stony asteroids and metallic asteroids. Considering all models, impactor diameter is in the range of 5.1 km and 80.9 km (figure 3 and Table 1).

### 5 Estimation of the concentration of iridium in the K/Pg layer

We can calculate the expected concentration of iridium using the estimated mass of the impactor and the iridium concentrations measured in meteorites (Nichiporuk and Brown, 1962; Fischer-Gödde et al., 2010). For iridium concentrations in comets, we considered that the dust to volatiles ratio is 1:1 (Greenberg, 1998), so the iridium concentration in comets is the half of the concentration in chondritic meteorite. Taking K/Pg layer with a thickness of 5 cm (Smit and Hertogen, 1980) our models predict iridium concentration between 1 and 5 041 ppb (see Table 2). We can compare with the measured concentration in the layer of 25 ppb (Smit and Hertogen, 1980). In the table 2 we see that models that predict this concentration are M3 Stony and M2, M3 and M4 comet (rows in Bold Italics characters). The measured value of 25 ppb was found, in all cases, close to the lower limit of the calculated intervals. This implies that the mass of the impactor is on the lower limits of our calculated ranges; which in turn implies high speeds of the impactor, i.e., we would be talking about a fast asteroid or a long-period comet.

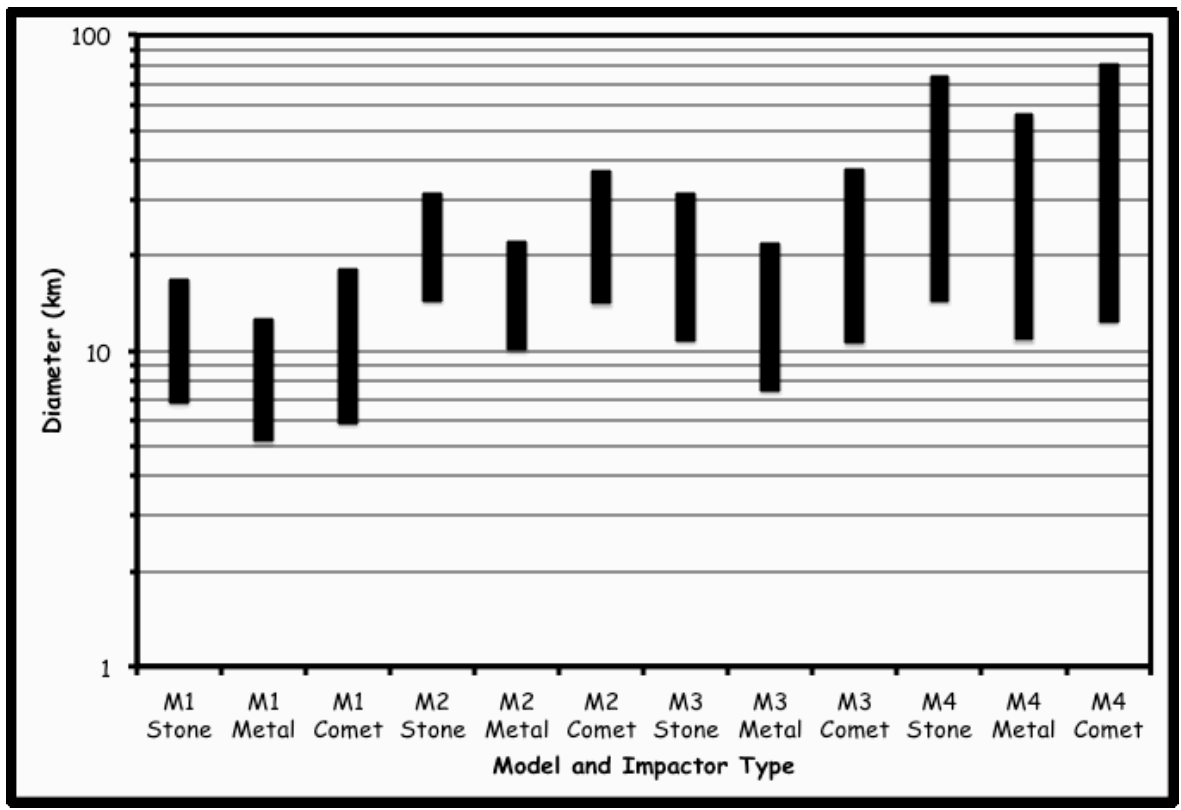

Figure 3. Diameter of the impactor according to the four models mentioned in text.

6 Discussion and conclusions

The first three models are empirical models developed from nuclear explosions or laboratory experiments, with energy less than or of the order of $10^{15}$ J. But the energy released in an impact, like the one in Chicxulub, is several orders of magnitude higher. For this reason, it is not clear that is valid to extrapolate the empirical models to those energies. For this, we needed to have a theoretical model to compare and to decide whether the extrapolation was physically correct, so we developed the fourth model. Although the two limits, upper and lower of the model, are both unrealistic, they allow us to say that the real energy lies between them, so we take the average of these two limits as our estimate of the energy.

Table 2
Estimated concentration of Ir in the K/Pg layer for different types of impactors. Rows in Bold Italic show valid cases (see text).

| Model | Type of Impactor | Ir (ppb) | Total Mass of Ir in the Impactor (kg) | | Concentration of Ir in the K/Pg layer (ppb) | |
|---|---|---|---|---|---|---|
| | | | Min | Max | Min | Max |
| 1 | Asteroid | 472$^a$ | 2.7x10$^8$ | 4.0x10$^9$ | 4 | 58 |
| | Comet | 236$^b$ | 4.2x10$^7$ | 1.3x10$^9$ | 1 | 18 |
| 2 | Stony | 472$^a$ | 2.5x10$^9$ | 2.7x10$^{10}$ | 36 | 385 |
| | Metallic | 3700$^c$ | 1.6x10$^{10}$ | 1.7x10$^{11}$ | 229 | 2428 |
| | Comet | 236$^b$ | 5.7x10$^8$ | 1.1x10$^{10}$ | 8 | 155 |
| 3 | Stony | 472$^a$ | 1.1x10$^9$ | 2.7x10$^{10}$ | 15 | 385 |
| | Metallic | 3700$^c$ | 6.5x10$^9$ | 1.6x10$^{11}$ | 94 | 2375 |
| | Comet | 236$^b$ | 2.4x10$^8$ | 1.1x10$^{10}$ | 3 | 158 |
| 4 | Asteroid | 472$^a$ | 2.5x10$^9$ | 3.5x10$^{11}$ | 36 | 5041 |
| | Comet | 236$^b$ | 3.7x10$^8$ | 1.1x10$^{11}$ | 6 | 1563 |

$^a$ (Fischer-Gödde, et al., 2010); $^b$ (Greenberg, 1998); $^c$ (Nichiporuk and Brown, 1962).

Energy estimations, resulting from the model one, are below the lower limit of model four. This leads us to think that in this case the extrapolation is not valid because it falls below the theoretical model. Therefore we only consider as valid estimations of energy, the results of the rest of the models (Figure 1).

The mass and diameter are calculated using estimated values of energy; therefore, we only considered these two parameters for the valid models mentioned in the previous paragraph.

Paquay et al., (2008) based on osmium isotope ratio $^{187}Os/^{188}Os$ propose a model to determine the size of the impactor. However, this method has great uncertainties because small variations in the ratio imply great changes in the size for projectile diameters greater than two kilometers. Our method does not have this uncertainty and can be used in the region in which the osmium method fails, which is precisely the case of the Chicxulub impactor. Taking all this into account, the estimated values of the energy, mass and diameter, can be seen in Table 1.

Shukolyukov and Lugmair (1998) analyzed samples from the Cretaceous to Tertiary boundary in Denmark and Spain and found that the isotopic composition of chromium is different from terrestrial samples and consistent with a carbonaceous chondritic impactor. On the other hand it was found that the composition of comets has a carbonaceous chondritic component (Jessberger et al. 1989) apart from the ice. Therefore, isotopic analysis cannot distinguish between comets and carbonaceous chondritic asteroids.

From our calculations of the iridium concentration we obtain that models that predict the observed concentration are M3 Stony and M2, M3 and M4 comet. The possibility of a metallic asteroid was eliminated. From this we concluded that the most probable impactor was a fast asteroid or a long-period comet with energy between $1.3 \times 10^{24}$ J and $5.8 \times 10^{25}$ J, mass between $1.0 \times 10^{15}$ kg and $4.6 \times 10^{17}$ kg, and diameter between 10.6 km and 80.9 km. Some authors have proposed a smaller size (~ 5.7 km) (Moore et al., 2013) but our model does not support that idea.